\documentclass[aps,prl,superscriptaddress,twocolumn,tightenlines]{revtex4}

\usepackage {graphicx}
\usepackage {amsmath}
\usepackage {amsfonts}
\usepackage {amssymb}
\usepackage {mathrsfs}
\usepackage {bm}

\usepackage {epstopdf}

\newcommand {\be}{\begin{eqnarray}}
\newcommand {\ee}{\end{eqnarray}}
\newcommand {\rmd} {{\rm d}}  
\newcommand{\dbar}{{\rm d}\mkern-6mu\mathchar'26}  
\newcommand{\diag}{{\rm diag}}

\begin{document}

\title {Rough edges in quantum transport of Dirac particles}
\author {Vladimir Cvetkovic}
\affiliation {Department of Physics \& Astronomy, The Johns Hopkins University,
Baltimore, MD 21218}
\author {Zlatko Tesanovic}
\affiliation {Department of Physics \& Astronomy, The Johns Hopkins University,
Baltimore, MD 21218}

\date {\today}
\begin {abstract}
We consider Dirac particles confined to a thin strip, 
e.g., graphene nanoribbon, with {\em rough} edges. The confinement
is implemented by a large mass
in the Hamiltonian or by imposing boundary conditions 
directly on the graphene wave-functions. The scattering
of a rough edge leads to a transverse channel-mixing and provides
crucial limitation to the quantum transport in narrow ribbons. 
We solve the problem perturbatively
and find the edge scattering contribution to the conductivity, which can be
measured experimentally. The case of Schr\"odinger particles
in a strip is also addressed, and the
comparison between Schr\"odinger and Dirac transport is made.
Anomalies associated with quasi-one dimensionality, 
such as Van Hove singularities and localization, are
discussed. The violation of the Matthiessen rule is pointed out.
\end {abstract}


\maketitle

Recent discovery of graphene \cite {NovoselovGeim, YZhang}
has spurred much interest.
The honeycomb structure of the graphene lattice implies that carbon
orbitals are $sp^2$ hybridized, leaving one 
free ($p_z$) electron orbital per atom. Unlike the 
majority of systems in condensed matter physics,
at low energies these electrons effectively obey Dirac equations of
motion for massless fermions \cite {BreyFertig}, with the Fermi 
velocity being the speed  of light.  Another
example is a d-wave
superconductor whose electrons behave relativistically in the vicinity of
nodal points \cite {ZlatkoQED}. 
The ``relativistic'' character of graphene
implies that its properties should be essentially different when
compared to more traditional systems,
whose electrons obey Schr\"odinger equation \cite {GeimReview}.

The central problem addressed in this Letter concerns transport properties
of narrow graphene ribbons (strips).
In general, the transport in thin and narrow structures is limited  
due to the restrictions imposed
by quantum mechanics \cite {TJMfilm}. 
Classically, a beam of particles may be collimated as
finely as desired so that it does not interact 
with the boundaries of a sample.
As a consequence -- in absence of ordinary impurity
scattering -- the mean free path of an electron
is infinite, leading to vanishing resistivity. 
Due to the quantum mechanical uncertainty, however,
a beam of particles must have a finite width: as the
lateral dimensions of a sample shrink toward and below this width,
the transport exhibits a crossover from the quasiclassical to the 
purely quantum
regime, governed by boundary scattering \cite {TJMfilm}. 
This is true both for Schr\"odinger and
Dirac particles. We expect our results to have broad 
implications for graphene based electronics. 

We use the Kubo formalism \cite {Mahan} to derive the expression for the
conductivity. Similar approach was used in Refs.\ \cite {LudwigFSG,
GusyninS, PeresGCN, AleinerEfetov, Ziegler}, although these authors were
interested in other issues. Alternatively, the conductivity can be
derived from the dielectric function \cite {HwangADS}, or equivalently \cite {Kubo2Landauer}
from the Landauer formula \cite {Tworzydlo, Bardarson,
Titov, KatsnelsonZitter, RyuMFL, Schomerus}. The quantum limitations to the transport properties
of graphene strips with {\em smooth} edges were considered in Ref.\
\cite {Tworzydlo}; obviously, this is
a different problem, exploring {\em ballistic} transport, 
in contrast to the present paper, where
the electrons are subject to a weak random scattering (impurities, phonons, etc.),
and have a {\em finite} mean free path. The transverse channel mixing due to
rough edge scattering in graphene junctions was numerically examined
in Ref.\ \cite {Schomerus}, however, this study too is centered on ballistic transport.

The confinement of Dirac fermions is a 
subtle issue due to the Klein paradox
\cite {KatsnelsonKlein}. This problem can be circumvented 
by using a large mass term
outside the strip in lieu of a static potential \cite {BerryM}. 
We employ this general method and refer to 
strips with such boundaries
as the Dirac strips or ribbons. The boundary 
conditions in graphene are
sensitive to the details of the
edge \cite {BreyFertig}, and can also be formulated on the microscopic
level \cite {Akhmerov}. For definiteness, we consider
the special case of a graphene strip with {\em metallic}
armchair edges, i.e., armchair strips/ribbons.
and assume the variations in its width
do not change the metallicity at the boundary \cite {BreyFertig,Blanter}.
The spectrum of metallic armchair strips 
is exceptional since it contains a band with zero transverse
momentum. Consequently, the armchair strips are anomalously good conductors,
especially in the vicinity of the Dirac point, 
where all the other boundary conditions
yield a finite gap. At a finite chemical potential 
these differences subside
but the aforementioned zero-band remains special as it does not get
scattered from the edges \cite{zigzag}. 
Our boundary conditions assume ``clean'' edges, i.e., the dangling
carbon bonds at the edges are inert and do not absorb environmental impurities
that might cause the current redistribution \cite {Zarbo} 
or localization of charge carriers \cite {LiLu}.

\begin{figure}
\includegraphics[width=0.45\textwidth]{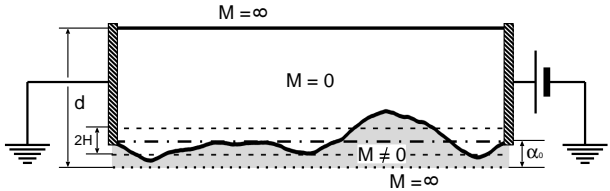}
\caption{A  thin strip of width $d$, whose bottom edge is
rough (the thick lines represent the physical edges). A buffer
layer (light gray) with large, but finite mass is attached to the rough edge
of the sample.}
\vskip -0.2truein
\label {FigBuffer}
\end{figure}

The scattering matrix due to rough edges is \cite {TJMfilm}
\be
  \hat V \propto \lbrack \lambda (\hat x) \lbrace \hat y, \hat p_y \rbrace,
  \hat H_y \rbrack, \label {VZlatko}
\ee
where $x$ and $y$ are the propagating and the transverse 
direction, respectively,
$\hat H_y$ is the transverse part of the Dirac Hamiltonian (straight edge),
and $\lambda (x)$ parametrizes the width profile. In contrast to the
Schr\" odinger case, the matrix Eq.\ \eqref {VZlatko} is {\em not} separable.
The matrix elements for a single rough edge can 
be found in Ref.\ \cite {KatsnelsonRibbons}, and 
it appears that the only way to proceed requires
inverting dense matrices of $n_c \times n_c$ size, with $n_c$ being the number
of transverse channels.
To overcome this sensitivity of the Dirac fermions 
to rough edges, we implement varying width boundary 
through an alternative construction, which mimics
the confinement of the Dirac fermions.
In place of an infinite mass jump at the edge, we allow the mass parameter
to change in two steps. At the edge the mass becomes large but finite,
followed by the infinite mass jump (Fig.\ \ref {FigBuffer}). This buffer
layer of variable width suppresses the wave-function
outside the strip while it simultaneously allows us
to recast the equivalent of the scattering 
matrix Eq.\ \eqref {VZlatko} in a
{\em quasi-separable} form, amenable to analytic treatment.

The Letter is organized as follows:
first, we consider Dirac states of a thin strip, and 
derive the Kubo formula for these Dirac states.
The Kubo formula is tested on a case of a {\em smooth} finite-width strip,
whose resistivity is needed in what follows.
Next, the buffer layer is introduced and the corresponding
matrix elements that mix transverse channels are found. The exact
mass parameter appearing in the matrix elements is determined by demanding
that the first-order perturbative corrections to the conductivity from a buffer
layer with a straight edge are in agreement with the physical result.
The effects of rough edges appear in the second-order perturbative corrections
to the energy. This leads to our main result, the 
conductivity due to rough edge scattering. 
Finally, we find the conductivity of
a thin strip whose electrons 
obey the Schr\"odinger equation \cite {TJMfilm}.
The comparison is made between the conductivities  of two
identical strips, one containing Dirac, and the other Schr\"odinger particles.

The 4-component Dirac equation
\be
  (v ~ \gamma_0 {\bm \gamma} \cdot {\bf p} - \mu) \Psi = \epsilon \Psi, \label {DiracEq}
\ee
with $\gamma_0 = \diag (\sigma_3, \sigma_3)$, $\gamma_1 = \diag (-i \sigma_2, i \sigma_2)$,
and $\gamma_2 = \diag (i \sigma_1, i \sigma_1)$, has solutions in form of plane waves
\be
  \Psi_{k m \xi, a} ({\bf r}) = e^{i k x} \left \lbrack u_{k, m, \xi, a}^{(+)}
  e^{i q_m y} + u_{k, m, \xi, a}^{(-)} e^{-i q_m y} \right \rbrack. \label {Psi0}
\ee
In our notation,  $k$ is the momentum
in the propagating direction, and $q_m$ is the transverse momentum. 
The boundary
conditions discretize the latter, yielding $q_m = (m + 1/2) \pi / d$, and $q_m = m \pi / d$
in the case of a Dirac, and an armchair strip
respectively. $m$ is taken nonnegative. $\xi$ is $\pm 1$ corresponding to
conductance and valence Dirac cones.
The elements of spinors $u_{k, m, \xi, a}^{\pm}$
are determined through the boundary conditions \cite {BreyFertig} and the
Dirac equation Eq.\ \eqref {DiracEq}.  Index $a=1, 2$ denotes two decoupled 
valleys for a Dirac strip, and two orthogonal states which admix the valleys
for an armchair strip. The energy of state Eq.\
\eqref {Psi0} is $\epsilon_{k m \xi} = - \mu + \xi v \sqrt {k^2 + q_m^2}$, with $\mu$ being
the chemical potential (equivalent to the gate voltage). The unperturbed
propagator is given by ${\cal G}^{m \xi a}_{n \zeta b} (i \omega_n; k) =
 {\cal G}^{(0)}_{m \xi a} (i \omega_n; k) {\delta_{m n} \delta_{\xi \zeta} \delta_{ab}} =
{\delta_{m n} \delta_{\xi \zeta} \delta_{ab}}/ ({i \omega_n - \epsilon_{k m \xi}})$.

The (longitudinal) current operator for a Dirac particle is $j_x ({\bf r}) =
e v \Psi ({\bf r})^\dagger \gamma_0 \gamma_1 \Psi ({\bf r})$. Starting with the Kubo formula,
and invoking the Lehmann representation for the propagtors \cite {Mahan},
we find the one-loop contribution to the dc conductivity for Dirac particles
\be
  \sigma_{dc} &=& \frac {e^2 v^2}{2 \pi d} \int \dbar k \sum_{m \zeta \xi a b} \sum_{n \eta \rho c d}
  {\cal A}^{n \eta c}_{m \zeta a} (0, k) {\cal A}^{m \xi b}_{n \rho d} (0, k) \times \nonumber \\
  && \frac {
  \left ( k \sigma_3 + q_m \sigma_2 \right )_{\zeta \xi}
  \left ( k \sigma_3 + q_n \sigma_2 \right )_{\eta \rho}  }
  {\sqrt {(k^2 + q_m^2) (k^2 + q_n^2)}}. \label {KuboAA}
\ee
Here, ${\cal A}^{m \xi a}_{n \zeta b} (E, k) =
- 2 ~ {\rm Im} ~ {\cal G}^{m \xi a}_{n \zeta b} (E + i \delta; k)$
is the spectral function matrix.

The Kubo formula Eq.\ \eqref {KuboAA} reproduces the Dirac point conductivity,
both the Gaussian \cite {LudwigFSG} and the universal values \cite {Fradkin86},
depending on the spectral function used. We use the Eq.\ \eqref {KuboAA}
to find the conductivity of a straight strip of width $d$,
where particles weakly scatter with an average lifetime
$\tau = 1 / (2 \Gamma )$. A straightforward calculation yields
\be
  \sigma_0 = \frac {e^2}{2 \pi} \frac {4 v}{d \Gamma} {\sum_m}^* \sqrt {1 - Y_m^2}, \label {sigma0}
\ee
where $Y_m = v q_m / | \mu |$, and the star indicates summation over all $m$-s
with $Y_m < 1$.
When a strip is wide (or $\mu$ large), the conductivity converges to
$\sigma = \frac {e^2}{ 2 \pi} (\mu / \Gamma)$ regardless the edge type.

We now implement the confinement of the Dirac particles by means of a
{\em finite} mass buffer layer. The basic idea is that
in a lattice system, it suffices
for the wave-function to fall-off faster than the lattice spacing to
effectively prohibit hopping
between the sites at the edge and their (non-existing) neighbours.
Hence, mass $M$ has to be large, and its 
precise value will be determined shortly. Since 
the buffer layer serves as
an extension of the strip exterior, this mass 
should be of the same type as
the infinite mass confining the 
fermions to the strip \cite {CommentMass}.
For a Dirac strip that is the {\em chiral} mass, and
a buffer layer of variable width $\alpha (x)$
introduces a perturbation 
$\hat V = M v^2 \gamma_0$, for $0 \le y \le \alpha (x)$.
The armchair boundary conditions are instead recreated via a
$\gamma_3
= \left ( \begin {array}{c c} 0 & -i \sigma_2 \\ -i \sigma_2 &0 \end {array} \right )$
``mass term''; the buffer layer perturbation for this edge is
$\hat V = M v^2 \gamma_0 \gamma_3$.
Using $t = - i \gamma_2$, 
the matrix elements to the leading  order in $\alpha$ are 
\be
  V^{l n \zeta a}_{k m \xi b} &=& \frac {M v^2}{4 d} \widetilde {\left \lbrack
  \alpha (x)^2 \right \rbrack}_{k-l}
  \left \lbrack (f_{l n \zeta a}^*)^\nu t_{\nu \mu}
  (f_{k m \xi b})^\mu \right \rbrack, \label {Vmatrix}
\ee
where summation over $\mu, \nu$ is assumed, and $\widetilde {\lbrack {\alpha} (x)^2 \rbrack}$
is the Fourier transform of $ {\alpha} (x)^2$. Defining $z_{k m \xi} = \xi ( k + i q_m)
/ \sqrt {k^2 + q_m^2}$, we have
$f^D_{k m \xi 1} = (- i q_m (1 + z_{k m \xi}), 1 - z_{k m \xi}, 0, 0)^T$,
and $f^D_{k m \xi 2} = (0, 0, - i q_m (1 - z_{k m \xi}^*), (1 + z_{k m \xi}^*) )^T$ for Dirac strips,
while for armchair strips  $f^A_{k m \xi 1} = (- i q_m z_{k m \xi}, 1, i q_m, z_{m k \xi})^T$,
and $f^A_{k m \xi 2} = (f^A_{k m \xi 1})^*$.
Importantly, the channel mixing 
matrix Eq.\ \eqref {Vmatrix} is quasi-separable.

Now, consider a buffer layer of a uniform and
small width $\alpha_0$. Both
$M$ and $\alpha_0$ enter Eq.\ \eqref {Vmatrix} as free parameters.
It is, however, clear that the first order corrections to the conductivity will
be proportional to $M \alpha_0^2$, while, according to Eq.\ \eqref {sigma0},
they should be proportional
to $\alpha_0$. The conclusion is that the product $M \alpha_0$ 
has to be a well defined constant.
This is rather natural: as the buffer layer shrinks, the suppression of the
wave function, driven by the inverse mass, is enhanced.
To find this constant, we recall that the change of
the conductivity by an infinitesimally thin buffer layer should satisfy:
\be
  \delta \sigma = \sigma (d) - \sigma (d - \alpha) \approx \alpha ~ \rmd \sigma_0 (d) / \rmd d~.
  \label {find_g}
\ee
The right hand side of Eq.\ \eqref {find_g} follows from 
Eq.\ \eqref {sigma0}. The left hand
side can be found from Dyson equation
\be
  {\cal G}^{n \zeta a}_{m \xi b} &=& {\cal G}^{(0)}_{n \zeta a} \delta_{nm} \delta_{\zeta \xi}
  \delta_{ab} + \nonumber \\
  &&{\cal G}^{(0)}_{n \zeta a} (1 - V^{k n \zeta a}_{k s \rho c} {\cal G}^{(0)}_{s \rho c})^{-1}
  V^{k s \rho c}_{k m \xi b} {\cal G}^{(0)}_{m \xi b}, \label {DysonV}
\ee
where argument $(i \omega_n; k)$ is assumed. The self-energy 
following from Eq.\ \eqref {DysonV}
is real, yielding the ``effective'' dispersion 
$\epsilon_{k m \xi}' = \epsilon_{k m \xi} + \xi M v^2
\alpha_0^2 q_m^2 / (d \sqrt{k^2 +q_m^2} )$ for both Dirac and
armchair strips.  
The corrections to the conductivity caused by
this dispersion are now substituted in \eqref {find_g}, 
implying that the mass in \eqref {Vmatrix} has the 
channel dependent form
$M = 1 / (v \alpha_0 Y_m Y_n)$. Thus, 
an infinitesimally thin buffer layer
perturbatively -- at the lowest order -- reproduces
the physical behavior expected from a layer of infinite mass.
A {\em metallic} armchair ribbon needs
special attention: the coupling for 
scattering into its {\em lowest} energy channel
($Y_0 = 0$) appears infinite, thus rendering the perturbative approach invalid.
This is reflective of an anomalous character of such 
a state, stemming from special boundary conditions.
Since this state is impervious to the mass term, we
ask how it is affected by the 
change of the strip width
{\em exactly}, i.e., we directly
examine the matrix elements $V_{0 m}$, Eq.\ \eqref {VZlatko}. 
They all
vanish and hereinafter we ignore the scattering to/from this channel.

We are now ready to find the second-order 
self-energy. Due to the quasi-separability
of the scattering potential Eq.\ \eqref {Vmatrix}, we use the following ansatz
$W_{n \zeta a, m \xi b} (i \omega_n; k) = (f_{k n \zeta a}^*)^\nu \hat W_{\nu \mu} (i \omega_n; k)
(f_{k m \xi b})^\mu$. The self energy and the propagator equations are cast as
\be
  \hat W (i \omega_n; k)
  &=& (\frac {M v^2 \alpha_0}{2 d})^2 \int \dbar l ~
  {\langle \widetilde {w}_{k - l} \widetilde {w}_{l - k} \rangle} \times 
  \label {Wselfc} \\ 
  &&t g^{(0)} (i \omega_n; l) \lbrack \hat 1 - \hat W (i \omega_n)
  {g}^{(0)} (i \omega_n; l) \rbrack^{-1} t, \nonumber \\
  {\cal G}^{n \zeta a}_{m \xi b} &=&  {\cal G}^{(0)}_{n \zeta a}
  \delta_{nm} \delta_{\zeta \xi} \delta_{ab} + \label {Gselfc} \\
  &&{\cal G}^{(0)}_{n \zeta a} (f_{n \zeta a})^\dagger \lbrack \hat 1 - \hat W
  {g}^{(0)} \rbrack^{-1} \hat W (f_{m \xi b})
  {\cal G}^{(0)}_{m \xi b}. \nonumber
\ee
The projected propagator is defined as
\be
  g^{(0) \nu \mu} (i \omega_n; k) = 
  \sum_{m \xi a} (f_{k m \xi a})^\nu {\cal G}^{(0)}_{m \xi a}
  (i \omega_n; k) (f_{k m \xi a}^*)^\mu. \label {Gproj}
\ee

In \eqref{Wselfc}, $w(x)$ is the deviation from 
the average buffer thickness
$\alpha_0$. For simplicity, 
we assume the white noise
edge profile characterized by $\langle w_k w_{-k} \rangle = a H^2$, 
where $a$ is the lattice spacing,
and $H^2$ measures the rms fluctuations in the strip width. Combining
the self-energy
from Eqs.\ (\ref{Wselfc}, \ref{Gselfc}) and Kubo formula 
\eqref {KuboAA}, yields the leading order conductivity 
expressed universally as
\be
  \sigma' = \frac {e^2}{2 \pi} \frac {4 v^2 d}{\mu^2 a H^2} {\sum_{m}}^*
  \sqrt {1 - {Y_m^2}} \times \qquad \qquad \qquad \qquad \qquad  \label {sigma_big}  \\
  \left \lbrace {\sum_n}^* \left ( \frac {2 + \frac 1{Y_m^2} + \frac 1{Y_n^2}}
  {\sqrt {1 - Y_n^2}} -
  \sqrt {1 - Y_m^2} \left (\frac 1 {Y_n^2} + \frac 1 {Y_m^2} \right ) \right ) \right \rbrace ^{-1}.
  \nonumber
\ee
This is the main result of this Letter.

\begin{figure} 
\includegraphics[width=0.24\textwidth]{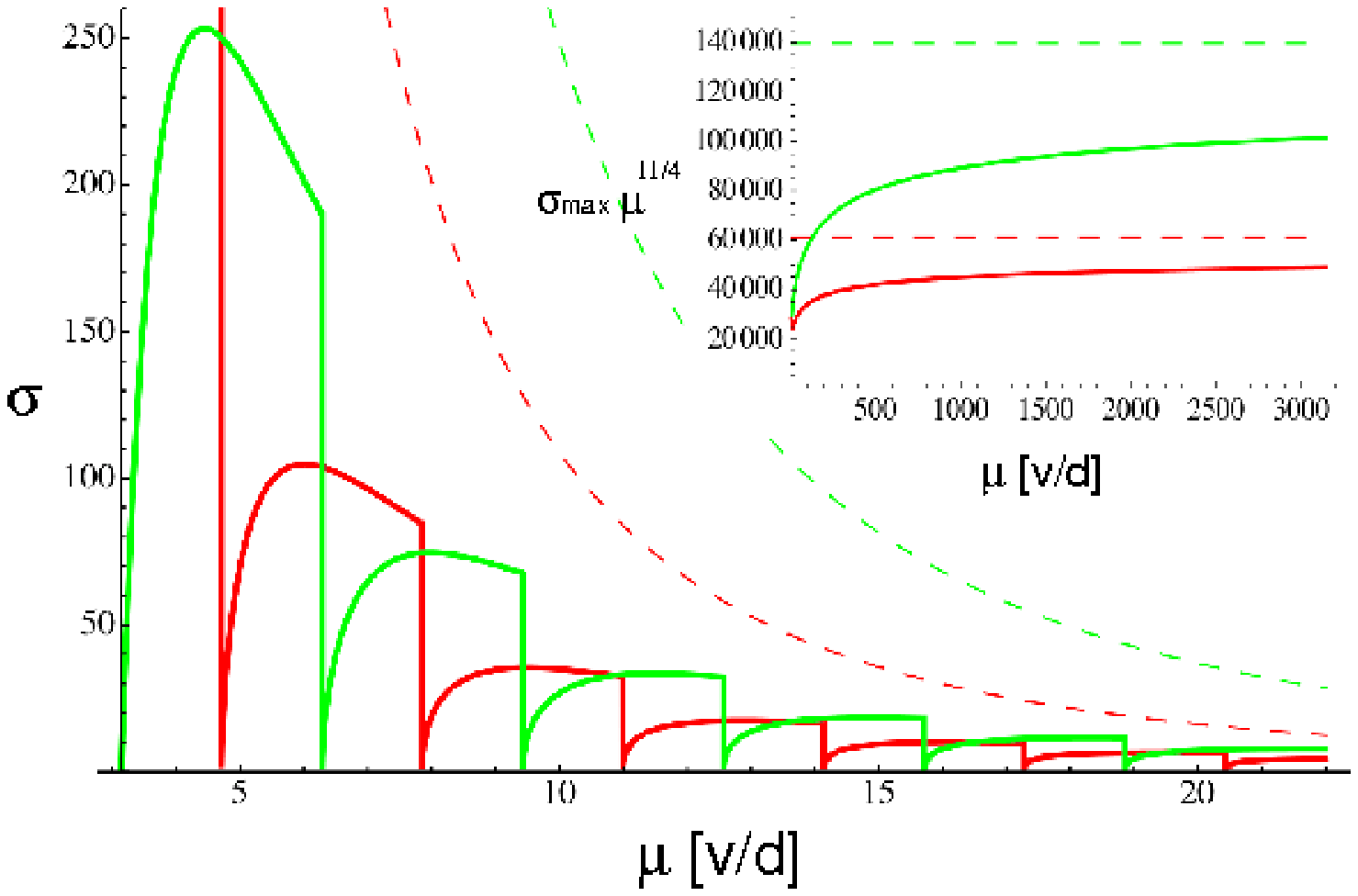}%
\includegraphics[width=0.24\textwidth]{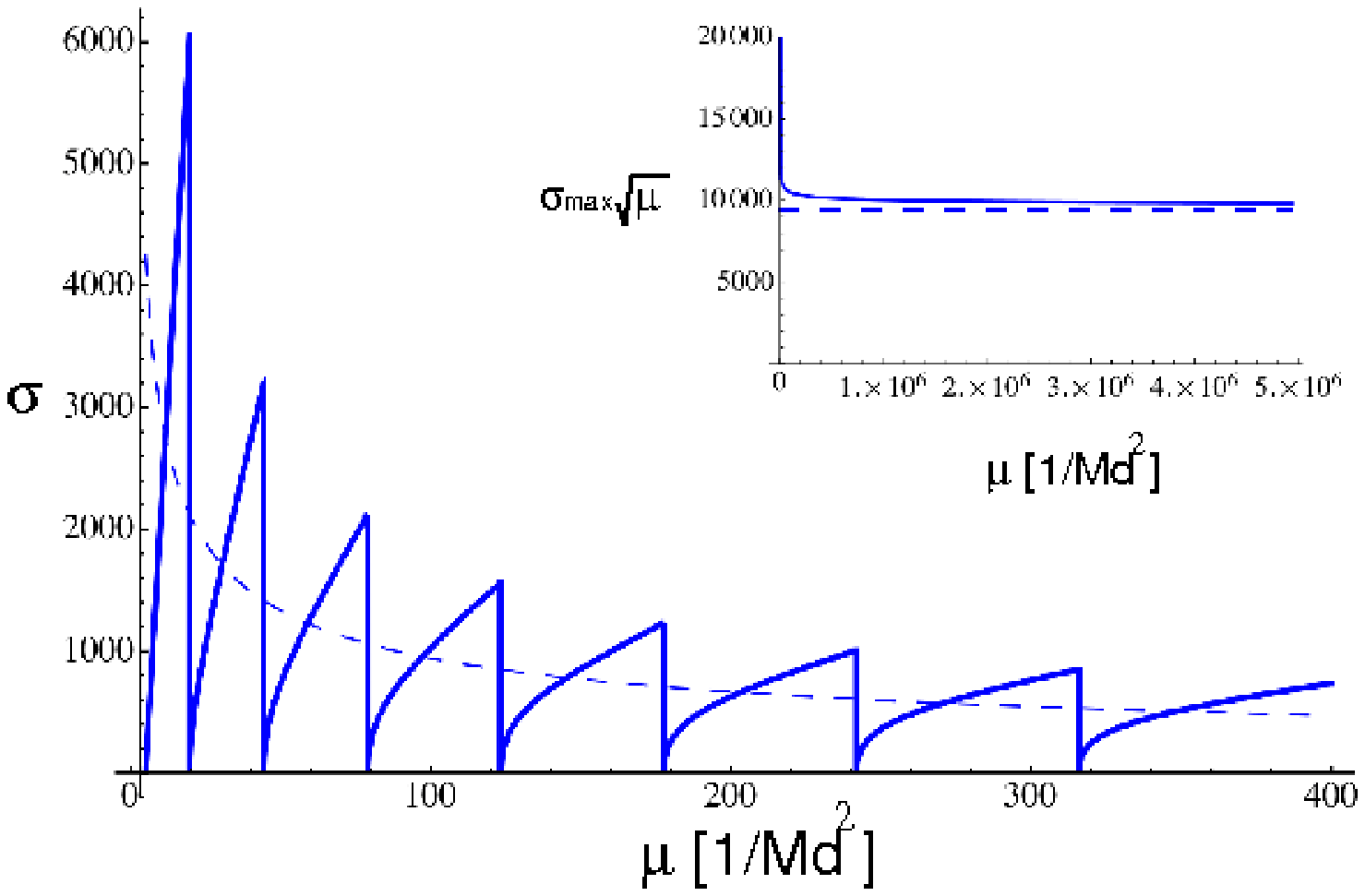}
\begin {center} a) \hskip 0.22\textwidth b)  \end {center}
\caption{The conductivities of a thin strip with a rough edge in units of $e^2 / h$. We set
$d = 1$, and $v = 1$ (Dirac, a), or $M=1$  (Schr\"odinger, b). The roughness
is $d^3 / a H^2 = 10^4$. In all three cases,
the Van Hove singularities appear; they are equidistant for
Dirac particles, while  their distance progressively grows
in the Schr\"odinger case. The conductivity(ies) of the Dirac particles
falls off considerably faster ($\mu^{-11/4}$) than that of 
the Schr\"odinger ones ($\mu^{-1/2}$). The inset
shows the asymptotics: the dashed lines correspond to
to Eqs.\ \eqref {sigma_max} and \eqref {sigma_max_S}. The conductivity
of an armchair strip (a; green) is on average greater by factor $2.28$ than
that of a Dirac strip (a; red).}
\vskip -0.1truein
\label {Fig_sigma_big}
\end{figure}

Unfortunately,
the summation in Eq.\ \eqref {sigma_big} is difficult to carry
out analytically and must instead be performed
numerically for each $\mu$ and $d$. 
An example is plotted in Fig.\ \ref {Fig_sigma_big}a.
Note sharp
drops (to zero) of the conductivity each time a new channel is open. With 
$\mu$ at the bottom of a newly open band, the density 
of states experiences Van Hove singularity 
(absent in
higher dimensions). As the number of states available for scattering
increases, so does the self-energy, and the ``effective'' lifetime
of excitations goes to zero. In Eq.\ \eqref {sigma_big}, 
this corresponds to
the situation  when the highest $Y_m$ is close to one.

For a wide strip, or large $\mu$,
one can estimate the asymptotic behavior of the 
maxima in the conductivity Eq.\ 
\eqref {sigma_big}, i.e., its
values just before the opening of a new channel. 
For a Dirac strip, the maxima converge to
\be
  \sigma_{max, D} \cong \frac {e^2}{2 \pi} \frac {v^{11/4} d^{1/4}}{\mu^{11/4} a H^2} 4
  \left ( \frac {2 \pi}{\pi - 2} \right ) ^{1/4}.
  \label {sigma_max}
\ee
In the case of an armchair strip with the same width the result is $3^{3/4} \approx 2.28$ 
times larger. Although Fig.\ \ref{Fig_sigma_big}a shows approximately equal conductivity maxima
at low chemical potential, as $\mu$ increases, so does the armchair strip become
better conductor as compared to the Dirac strip.
The conductivity converges slowly to Eq.\ \eqref {sigma_max},
as shown in the inset of Fig.\ \ref {Fig_sigma_big}a.

For comparison, we consider the case of a strip with
``Schr\" odinger'' carriers. The Fermi energy is denoted
by $\mu$, the same as the chemical potential of the Dirac case. 
The dispersion in an unperturbed system is
$\epsilon_{k m} = - \mu + (k^2 + q_m^2) / (2 M)$, $M$ being the
electron mass here. The formalism of Ref. \cite {TJMfilm}
is readily adapted to a thin strip.
Assuming the same parameters as before, the conductivity is 
\be
  \sigma' = \frac {e^2}{2 \pi} \frac {d}{a H^2 M \mu} {\sum_m}^* \frac {\sqrt {1 - Y_m^2}} {Y_m^2}
  \left \lbrace {\sum_n}^* \frac {Y_n^2}{\sqrt {1- Y_n^2}} \right \rbrace^{-1}, \label {sigma_bigS}
\ee
and is plotted in Fig.\ \ref {Fig_sigma_big}b. 
As in the previous case,
$Y_m = q_m / \sqrt {2 M \mu}$. Van Hove singularities
are appearing here too; however, in contrast to the Dirac case,
the singularities are not equidistant, and are instead
distributed as squares of integers, due to
the parabolic dispersion. 
The other qualitative difference is reflected in 
the asymptotic trend for the conductivity maxima.
For wide strips, the conductivity peaks follow the envelope function
\be
  \sigma_{max, S} \cong \frac {e^2}{2 \pi} \frac {4 d^2}{3 a H^2 \sqrt {2 M \mu}}. \label {sigma_max_S}
\ee
The conductivity has a sharp $d^2$ dependence (compare to $d^3$ for thin films \cite {TJMfilm}).

There is a notable qualitative difference between the asymptotic behaviors
Eqs.\ \eqref {sigma_max_S} and \eqref {sigma_max}. For a Schr\"odinger
strip, the major contribution to the conductivity comes from the lowest
bands, as these states scatter the least from rough edges. Accordingly, the sum in Eq.\
\eqref {sigma_max_S} is asymptotically proportional to the number of open channels,
and the overall conductivity is proportional to $\mu^{-1/2}$.
The Dirac states in the lowest bands are, on the other hand, the most susceptible
to the edge scattering, and their contributions are largely suppressed. Hence,
the conductivity Eq.\ \eqref {sigma_max} decreases rapidly ($\sim \mu^{-11/4}$)
as the chemical potential increases. The high impact of the edge scattering on the
Dirac particles is also evident in the weak $d^{1/4}$ width dependence in Eq.\ \eqref {sigma_max}.
These peculiar power laws are the consequence of the conductivity-per-channel
function in Eq.\ \eqref {sigma_big}.

In quasi one-dimensional systems, 
one must be mindful of inevitable localization \cite {HalperinLoc}. Its effects are
discussed elsewhere \cite {CTnew}.
Generally, one anticipates a
certain localization length $L$
for electrons. The results presented here should be valid for weak disorder, i.e. 
whenever the strip is shorter than $L$.

In thin films, the scattering from a rough surface yields a channel dependent
mean free path, whereas the impurity scattering is channel independent, 
resulting in the violation of the Matthiessen rule \cite {Matthiessen}. 
In thin strips the situation is similar,  the channel mean free path $l_m'$ from
Eq.\ \eqref {sigma_big}
is  not proportional to
the impurity mean free path of a corresponding channel
$l^0_m = v \sqrt {1 - Y_m^2} / ( d \Gamma k_F )$, Eq.\ \eqref {sigma0},
hence the Matthiessen rule is violated here, too. 

In this Letter we have developed a perturbative approach to the problem of the
edge scattering in Dirac strips. The
particle confinement is implemented in the manner
that allows us to solve the problem analytically. 
The conductivity is found to the leading order in the edge roughness. We also analyze thin strips
with Schr\"odinger particles and compare the results to those
for Dirac strips.
In both cases, the conductivity develops 
Van Hove singularities; however,
different dispersions mean that 
the singularities for Dirac particles are
equidistant in the chemical potential, while those for the
Schr\"odinger ones are not. The higher sensitivity of Dirac particles to the
edge roughness suppresses the otherwise most conductive channels,
and accordingly,  the (average) conductivity fall off as a 
function of the chemical potential is much
faster as compared to the Schr\"odinger case.

This work was
supported in part by the NSF grant DMR-0531159.

\bibliographystyle{apsrev}

\begin{thebibliography}{10}
\bibitem {Blanter} If the metallicity changes along the strip, e.g., the width is changed by a 
single row of atoms, numerical studies show segmentation of charges,
and the tight binding model becomes a more appropriate representation:
I.\ Martin, Ya.\ M.\ Blanter, arxiv.org/0705.0532
\bibitem {CommentMass} More precisely, one should introduce both the chiral and PT masses
in the buffer layer. The condition Eq. \eqref {find_g} imposed on both valleys then
sets $M_{PT}$ to zero, and the chiral mass to the value given in the text.
\end{thebibliography}

\end {document}